\documentclass[12pt]{aastex}

\begin{document}

\title{First and second-type self-similar solutions of implosions and
explosions containing ultra-relativistic shocks} \author{Re'em Sari \\
Theoretical Astrophysics 130-33, California Institute of Technology,
\\ Pasadena, CA 91125, USA }

\begin{abstract}
We derive self similar solutions for ultra-relativistic shock waves
propagating into cold material of powerlaw density profile in radius
$\rho \propto r^{-k}$. We treat both implosions and explosions in
three geometries: planar, cylindrical and spherical. For spherical
explosions these are the first type solutions of Blandford and McKee
for $k<4$ and the second type solutions found by Best and Sari
for $k>5-\sqrt{3/4}$. In addition we find new, hollow (with evacuated
interior), first type solutions that may be applicable for
$4<k<17/4$. This ``sequence'' with increasing $k$ of first type
solutions, hollow first type solutions, and then second type solutions
is reminiscent of the non-relativistic sequence.  However, while in
the non relativistic case there is a range of $k$ which corresponds to
a ``gap'' - a range in $k$ with neither first nor second type solutions
which separates the hollow first type solutions and the second type
solutions, here there is an ``overlap'': a range of $k$ for which
current considerations allow for both hollow first and second type
solutions. Further understanding is needed to determine which of the
two solutions apply in this overlap regime. We provide similar
exploration for the other geometries and for imploding
configurations. Interestingly, we find a gap for imploding spherical
shocks and exploding planar shocks and an overlap for imploding planar
solutions. Cylindrical configurations have no hollow solutions and
exhibit direct transition from first type to second type solutions,
without a gap or an overlap region.
\end{abstract}

\maketitle

\section{Introduction}
Self similar solutions provide some of the greatest simplifications to
one dimensional flows. Self similarity allows the reduction of the
partial differential equations, which contain two independent
variables (space and time), into a set of ordinary differential
equations, where the single independent variable is a combination of
space and time.  The ODEs are then relatively easy to solve
numerically or even analytically in some cases. They describe the
asymptotic behavior of one dimensional flows in a variety of
circumstances, typically far away from the initial conditions and
provided that the boundary conditions contain no spatial scale (Some
exceptions apply. For example, self similarity can prevail in
exponential density gradient in planar geometry.).

Perhaps the most famous hydrodynamic self similar solution is that due
to Sedov\cite{sedov}, Von Neumann\cite{vonneumann} and
Taylor\cite{taylor} known as the Sedov-Taylor solution.  It describes
an explosion in which a strong shock wave propagates into cold
surroundings whose density profile decreases as\footnote{In the non
relativistic literature, the density powerlaw index is usually denoted
by $\omega$ while in the relativistic temperature the letter $k$ is
used. Here we use the letter $k$ for both as our focus is the
relativistic regime} $\rho \propto r^{-k}$. They used conservation of
energy to obtain the scaling of the shock radius as a function of
time.  Such solutions are called first type solutions.  Zeldovich and
Raizer \cite{ZR66} discuss a self similar solution describing
imploding shock waves in a constant density environment. In contrast
to the strong explosion problem, energy considerations cannot be used
to deduce the scaling of the shock radius as a function of
time. Instead the scaling of the radius as function of time must be
found by demanding that the solution pass through a singular point of
the equation. Such solutions are called self similar solutions of the
second type. More recently, Waxman and Shvarts\cite{WS93} showed that
if the density falls fast enough ($k>3$), energy considerations give
the wrong scaling. They also showed that the solution should be of the
second type for $k>3.26$.

Recent astrophysical discoveries, mostly the afterglows of gamma ray
bursts, led to an increased interest in the relativistic analogs of
these solutions. We find that there is a considerable similarity
between the relativistic and the the non relativistic regimes. The
relativistic version of the Sedov-Taylor solution was found by
Blandford and McKee in 1976 [BM hereafter, \cite{BM76}]. They provide
the solution to the ultra relativistic strong explosion problem in
spherical geometry for density profiles with $k<4$. Best and
Sari\cite{BS00} have found solutions of the second type for
$k>5-\sqrt{3/4}$. Perna and Vietry\cite{PV02} found relativistic
solutions for shock waves propagating in an exponential density
gradient. These are the relativistic analogs of the non relativistic
solutions first found by Raizer\cite{raizer}.

In this paper, we provide a more complete exploration of the possible
self similar solutions in the ultra-relativistic case. We provide
solutions in three geometries, planar, cylindrical and spherical for
both imploding and exploding shock waves.  Our investigation reveals
some unexpected puzzles.  For example, for spherical explosions, we
find a range of density profiles $5-\sqrt{3/4}<k<17/4$ for which
current considerations allow for both first and second type
solutions. Understanding which of these solutions will apply is left
for later research. Fortunately, all our solutions are analytic. This
is an advantage over the non relativistic case where the self similar
solutions are either implicit (first type) or numerical (second
type). This property may facilitate the understanding of the interplay
between first and second type solutions.

The plan of this paper is as follows. We begin in section \S2 by
writing down the one dimensional flow equations in planar, cylindrical
and spherical symmetries. We then take the ultra relativistic limit
and assume self similarity so the partial differential equations are
reduced into ordinary differential equations. In section \S3 we
explore the range in density powerlaw index for which first type
solutions, which obey global conservation law, apply and contain a
finite energy. In section \S4 we find second type solutions and
explore the regimes where those are valid.

\section{The self similar equations in arbitrary dimensions}
The flow equations are given by the conservation of energy, momentum,
and particle number. Since in the relativistic case the particle
number density does not play a role in the dynamics of the system, the
conservation of particles equation is decoupled from the other two
equations. Energy and momentum conservation yield
\begin{eqnarray}
\frac{\partial }{\partial t}\gamma ^{2}(e+\beta ^{2}p)+\frac{1}{r^{\alpha }}%
\frac{\partial }{\partial r}r^{\alpha }\gamma ^{2}\beta (e+p) &=&0 \\
\frac{\partial }{\partial t}\gamma ^{2}\beta (e+p)+\frac{1}{r^{\alpha }}%
\frac{\partial }{\partial r}r^{\alpha }\gamma ^{2}\beta ^{2}(e+p)+\frac{%
\partial }{\partial r}p &=&0,
\end{eqnarray}
and the particle conservation equation is 
\begin{equation}
\frac{\partial }{\partial t}\gamma n+\frac{1}{r^{\alpha }}\frac{\partial }{%
\partial r}r^{\alpha }\gamma \beta n=0,
\end{equation}
where $\alpha =0,1,2$ for planar, cylindrical and spherical symmetries
respectively.

We now assume that the flow has a characteristic Lorentz factor
$\Gamma$ and a related characteristic position $R$. $R$ evolves with
time in a way that describes motion with a Lorentz factor $\Gamma$. In
the extreme relativistic case this implies $\dot R=\sqrt{1-1/\Gamma^2}
\approx 1-1/2\Gamma^2$. For the problem of interest here, it is
natural to choose the position of the shock, and its Lorentz factor as
$R$ and $\Gamma$. The characteristic length scale of the flow behind
the shock must be $R/\Gamma ^{2}$, so the similarity variable is
\begin{equation}
\zeta =\frac{R-r}{R/\Gamma ^{2}}=\Gamma ^{2}(1-r/R).
\label{zeta}
\end{equation}
BM doubted the use of equation (\ref{zeta}) for $k>4$; however, Best
and Sari \cite{BS00} demonstrated that this definition of the self
similar variable is applicable in general.

Note that in our notation $\dot R$ is always positive.  For converging
solutions, this implies that $R$ is negative, growing towards zero. In
our notation, therefore, converging solutions have negative
$\zeta$. This choice avoids the need to keep a $\pm$ sign in the
expression of $\dot R$ for imploding and exploding solutions.

The self similar solution exhibits time dependent characteristic
pressure and density, which we denote as $P(t)$ and $N(t)$. For
consistency with the BM notation we define the self similar variables
as
\begin{eqnarray}
\gamma ^{2}(r,t) &=&\frac{1}{2}\Gamma ^{2}(t)g(\zeta ), \\
p(r,t) &=&P(t)f(\zeta ), \\
n(r,t) &=&N(t)h(\zeta)/g^{1/2}(\zeta)
\end{eqnarray}
where $g$, $f$ and $h$ are functions of the self similar variable
$\zeta$ and respectively describe the spatial profiles of the Lorentz
factor, pressure and density.

In terms of the self similar variables we have
\begin{equation}
\frac{\partial }{\partial t}=\dot{\Gamma}\frac{\partial }{\partial \Gamma }+%
\left[ 2\zeta \frac{\dot{\Gamma}}{\Gamma }+(\Gamma ^{2}-1/2)\frac{r}{R^{2}}%
\right] \frac{\partial }{\partial \zeta }+\dot{P}\frac{\partial }{\partial P},
\end{equation}
and
\begin{equation}
\frac{\partial }{\partial r}=-\Gamma ^{2}\frac{1}{R}\frac{\partial }{%
\partial \zeta }.
\end{equation}
The energy conservation equation now becomes:
\begin{equation}
0=2(-2m+\alpha -k)gf-(1+2(m+1)\zeta )(gf)^{\prime }+f^{\prime }
\end{equation}
As in the work of BM, we are motivated by this equation to define
\begin{equation}
\chi =1+2(m+1)\zeta,
\end{equation}
to obtain
\begin{equation}
0=2(-2m+\alpha -k)gf-2(m+1)\left[ \chi{d \over d\chi} (gf)-{d \over d\chi} f\right] .
\label{energyss}
\end{equation}
Here we have used the notation $\frac{t\dot{\Gamma}}{\Gamma }=-m/2$, $
\frac{t\dot{P}}{P}=-m-k$, and ${t\dot N \over N}=-m/2-k$. The equation
above shows that $m$ and $k$ must be constant to allow for a self
similar flow. The relation between the derivatives of $N$, $P$, and
$\Gamma$ follows from the boundary conditions of strong relativistic
shocks. Since we set $\chi=1$ at the shock, those boundary conditions
also imply $g(1)=f(1)=h(1)=1$ (see BM).

Repeating the same procedure for the momentum equation will again
result in equation (\ref{energyss}). This is because the energy and
momentum equations are identical to lowest order in $1/\Gamma$ and our
expansion discards any higher order terms. For this reason BM kept
higher order terms. Instead, we find it simpler to use the difference
equation between the energy and momentum conservation equations:
\begin{equation}
\frac{\partial }{\partial t}(2+\frac{1}{\gamma ^{2}})p+\frac{1}{r^{\alpha }}
\frac{\partial }{\partial r}r^{\alpha }(4-\frac{1}{\gamma ^{2}})p-2\frac{
\partial }{\partial r}p=0,
\label{eq:diffnoss}
\end{equation}
Keeping only lowest order terms is again sufficient here. With the
self similar variables equation \ref{eq:diffnoss} reads:
\begin{equation}
0=(m+1)
\left[ -\chi g^{2}{df\over d\chi }+4({df\over d\chi}g-{d g \over d\chi }f)\right]
+(-m-k+2\alpha )g^{2}f.
\label{eq:diff}
\end{equation}
Similarly, for the particle conservation equation we have:
\begin{equation}
{d\log h \over d\chi }=
(-kg^{2}-mg^{2}+\alpha g^{2}-2(m+1){dg\over d\chi })/(\chi
g^{2}-2g)/(m+1).
\label{eq:particle}
\end{equation}

We now solve equations (\ref{energyss}), (\ref{eq:diff}), and
(\ref{eq:particle}) as a set of linear equations to get $df/d\chi$,
$dg/d\chi$, and $dh/d\chi$ and obtain:
\begin{equation}
{1 \over g\chi}{d \log g \over d \log\chi }
=\frac{(7m+3k-2\alpha )-(m+\alpha )\chi g}
       {(m+1)(g^2\chi ^{2}-8g\chi +4)}.
\label{dgdchi}
\end{equation}
\begin{equation}
{1 \over g\chi }{d \log f\over  d\log \chi }=
\frac{4(2m-\alpha +k)-(m+k-2\alpha )g\chi }
       {(m+1)(g^2\chi ^{2}-8g\chi +4)}.
\label{dfdchi}
\end{equation}
\begin{equation}
{1 \over g\chi}{d \log h \over d\log\chi }=
\frac{2(9m+5k-4\alpha )-2(5m+4k-3\alpha)g\chi +(m+k-\alpha )g^{2}\chi ^{2}}
  {(m+1)(2-g\chi )(g^2\chi^2-8g\chi +4)}.
\label{dhdchi}
\end{equation}
For the case $\alpha =2$ these reduce to the equations given in BM.

\section{First type solutions}

The parameter $m$ in the equations above must be found in an
independent way. As is well known in the non-relativistic case, there
are two types of self-similar solution. In the first type, $m$ is
found by global conservation laws. For arbitrary geometry, $\alpha$,
in the ultra relativistic regime this requirement reads
\begin{equation}
E\sim \gamma ^{2}R^{-k}R^{1+\alpha }\sim R^{-m-k+1+\alpha }\sim const.
\end{equation}
Therefore first type solutions have
\begin{equation}
m=1+\alpha -k.
\end{equation}
Substituting this into the self similar equations and using the
boundary conditions $g(1)=f(1)=h(1)=1$ we obtain the very simple
solution:
\begin{equation}
g=\chi ^{-1}
\label{eq:gfir}
\end{equation}
\begin{equation}
f=\chi ^{\frac{4k-7-5\alpha }{3(2+\alpha -k)}}
\end{equation}
and
\begin{equation}
h=\chi ^{\frac{2k-3-2\alpha }{2+\alpha -k}}.
\end{equation}
This is a generalized form of the BM solution, and reduces to the BM
solution in spherical geometry where $\alpha =2$.

First type solutions can be valid only if they contain a finite amount
of energy. The energy in the solution is proportional to the integral
\begin{equation}
\int fg d\chi \sim \chi^{\frac{4k-7-5\alpha }{3(2+\alpha -k)}}.
\end{equation}
This is finite if $1<\chi$ and $\frac{4k-7-5\alpha}{3(2+\alpha -k)}<0$
or if $0<\chi<1$ and $\frac{4k-7-5\alpha}{3(2+\alpha -k)}>0$.

$1<\chi$ if $R(m+1)=R(2+\alpha-k)$ is positive. So the combination of
the two possibilities above requires $(4k-7-5\alpha)R<0$. Therefore,
explosions diverging to infinite distances ($R$ is positive) can be of
first type if $k<(7+5\alpha)/4$. Converging solutions (where $R$ is
negative approaching zero) are of first type only for
$k>(7+5\alpha)/4$.

For $\chi<1$ these solutions are hollow and end at $\chi=0$. The
Lorentz factor diverges at $\chi=0$, and it takes infinite time for a
fluid element to arrive from the shock ($\chi=1$) to $\chi=0$. Hollow
solutions exist for diverging shocks in spherical symmetry with
$4<k<17/4$ and for converging shocks in planar symmetry with
$2>k>7/4$.

\section{Second Type Solutions}

Second type solutions do not obey global conservation law. The true
problem, therefore, cannot be completely described by a second type
self similar solution. Those describe only part of the flow, in some
region of interest while other regions deviate from the solutions. In
order not to influence the self similar part, the non self similar
parts must be separated from it by a sonic point, where the equations
are singular. This requierment replaces the energy conservation as
means of deducing the scaling of Lorentz factor with radius,
i.e. finding $m$ (see \cite{WS93} for discussion of the non
relativistic case and \cite{BS00} for the relativistic case).

To find $m$ in second type solutions, we notice that the denominators
of equations \ref{dgdchi}, \ref{dfdchi}, and, \ref{dhdchi} is
independent of $\alpha$, therefore the sonic line has the same value
when expressed in terms of $g\chi $. The dependence of $m$ on $k$ can
be easily found, using the same method as in Best and Sari. Looking
for roots of the denominator of our equations we find the sonic line
to be
\begin{equation}
\label{gchisonic}
g\chi=4-2\sqrt{3}.
\end{equation}
While the fluid equations have other singular points\footnote{The
other singular points are $g\chi=2$, which represents a fluid element
maintaining a fixed $\chi$, and $g\chi=4+2\sqrt{3}$, which corresponds
to the negative characteristic having a fixed $\chi$. All those values
are independent of the geometry, i.e. independent of the parameter
$\alpha$.}, physical considerations (see ref. \cite{BS00}) show that
only the one given by (\ref{gchisonic}) is the one we are looking
for. This point separates the characteristic heading to the positive
direction from the shock, which also goes in the positive direction
per our definitions. Substituting this in the numerator of, for
example, the equation of $g$ and demanding that it vanish, we obtain:
\begin{equation}
m=-2\alpha (5-3\sqrt{3})+(3-2\sqrt{3})k.
\label{mII}
\end{equation}
This reproduces the results of Best and Sari for $\alpha=2$. The
hydrodynamic profiles can be found by substituting this value of $m$
in the equations for $g$, $f$ and $h$. The equations become very
simple for this value of $m$ since by definition the denominator and
the numerator have a common factor.

\begin{equation}
g=K\left[ \frac{1-\alpha }{-10\alpha +6\alpha \sqrt{3}+3k-2k\sqrt{3}+1}g\chi
-2(2+\sqrt{3})\right] ^{\left( 3-2\sqrt{3}\right) \left( k-3\alpha \right)
/\left( \alpha -1\right) }
\end{equation}
where $K$ is an arbitrary constant. The boundary conditions $g(1)=1$
implies
\begin{equation}
\label{eq:gsec}
g=\left[ \frac{g\chi (\alpha -1)+4\alpha (\sqrt{3}-1)-2k\sqrt{3}+4+2\sqrt{3}%
}{(\alpha -1)+4\alpha (\sqrt{3}-1)-2k\sqrt{3}+4+2\sqrt{3}}\right] ^{\left(
3-2\sqrt{3}\right) \left( k-3\alpha \right) /\left( \alpha -1\right) }.
\end{equation}

For the pressure, we get
\begin{equation}
\log f=K+\frac{(4-2\sqrt{3})(k-3\alpha )}{\alpha -1}\log \left[ g\chi(\alpha
-1)-4\alpha +4\alpha \sqrt{3}-2k\sqrt{3}+4+2\sqrt{3}\right]
\end{equation}
or, with the boundary conditions $f(1)=1$,
\begin{equation}
f=\left[ \frac{g\chi(\alpha -1)-4\alpha +4\alpha \sqrt{3}-2k\sqrt{3}+4+2\sqrt{3}%
}{\alpha -1-4\alpha +4\alpha \sqrt{3}-2k\sqrt{3}+4+2\sqrt{3}}\right] ^{(4-2%
\sqrt{3})(k-3\alpha )/\left( \alpha -1\right) }
\end{equation}

For the density we get:
\begin{equation}
\log h=K-
\]
\[
\frac{\left( 2\sqrt{3}-3\right) \left( 2k-1+\alpha \sqrt{3}
-3\alpha \right) \left( k-3\alpha \right) }{\left( -1-2\alpha \sqrt{3}+k
\sqrt{3}-\sqrt{3}+\alpha \right) \left( 1-\alpha \right) }\log \left[
(1-\alpha )g\chi +4\alpha -4\alpha \sqrt{3}+2k\sqrt{3}-4-2\sqrt{3}\right]
-
\]
\[ \frac{2\alpha -k-\alpha \sqrt{3}}{-1-2\alpha \sqrt{3}+k\sqrt{3}-
\sqrt{3}+\alpha }\log \left( g\chi -2\right).
\end{equation}
With the boundary conditions $h=1$ at $\chi =1$ we have:
\begin{equation}
h=\left[ \frac{(1-\alpha )g\chi +4\alpha -4\alpha \sqrt{3}+2k\sqrt{3}-4-2
\sqrt{3}}{(1-\alpha )+4\alpha -4\alpha \sqrt{3}+2k\sqrt{3}-4-2\sqrt{3}}
\right] ^{\frac{\left( 2\sqrt{3}-3\right) \left( 2k-1+\alpha \sqrt{3}
-3\alpha \right) \left( k-3\alpha \right) }{\left( \alpha -1-2\alpha \sqrt{3}
+k\sqrt{3}-\sqrt{3}\right) \left( \alpha -1\right) }}\left[ 2-g\chi \right]
^{\frac{-2\alpha +k+\alpha \sqrt{3}}{\alpha -1-2\alpha \sqrt{3}+k\sqrt{3}-
\sqrt{3}}}
\end{equation}

For cylindrical geometry, $\alpha =1$, the above form is invalid and
the solution (with the boundary condition of $g=1$ at $\chi =1$) reads
\begin{equation}
\log g=\frac{g\chi -1}{2(2+\sqrt{3})}.
\end{equation}
\begin{equation}
\log f={3-2\sqrt{3} \over 3} \left( g\chi-1 \right)
\end{equation}


The condition we used to demand a smooth transition through the sonic
point, which led to equation \ref{mII}, is necessary but not
sufficient. We still need to verify that the solution passes through
the sonic point in the relevant range of the independent variable
$\chi$. We therefore substitute $g\chi=4-2\sqrt 3$ in the expression
for $g$ and find the value $\chi_{sonic}$. This is displayed in the
figure below for the three possible values of $\alpha$. We demand that
$\chi_{sonic}$ is within the range of values that $\chi$ takes, i.e.,
we require $\chi_{sonic}>1$ for diverging solution with $m>-1$ or
converging solutions with $m<-1$, and we require $\chi_{sonic}<1$ for
diverging solutions with $m>-1$ or converging solutions with
$m<-1$. The results regarding the validity range of type-II solutions
are given in table I and shown schematically in figure 1.

\begin{table}[h]
\begin{center}
\begin{tabular}{|c|c|c|c|c|}
\hline
symmetry   & \multicolumn{2}{c|}{diverging solutions ($R \rightarrow \infty$)} & \multicolumn{2}{c|}{converging solutions ($R \rightarrow 0$)} \\
           & type-I     &                   Type-II       & Type-I            & Type-II   \\  \hline
planar     & $k<7/4$    &       $k>2$                     & $k>7/4$           & $k<1$                                    \\ \hline
cylindrical& $k<3$          &       $k>3$                     & $k>3$             & $k<3$                                    \\ \hline
spherical  & \,\,\,$k<17/4$\,\,\, &   $k>5-\sqrt{3/4} \cong 4.13$   & \,\,\,$k>17/4$\,\,\,          & $k<4$                              \\ \hline    
\end{tabular}
\caption{The validity range of second type solutions. The choice of
$m$ according to equation \ref{mII} guarantees that if a solution
passes through a sonic point, the solution is smooth there.  The
conditions listed in the table verify that the solution does in fact
pass though the sonic line within the physically relevant range of
$\chi$.}
\end{center}
\end{table}

\begin{figure}[ht!]
\epsscale{0.6}
\plotone{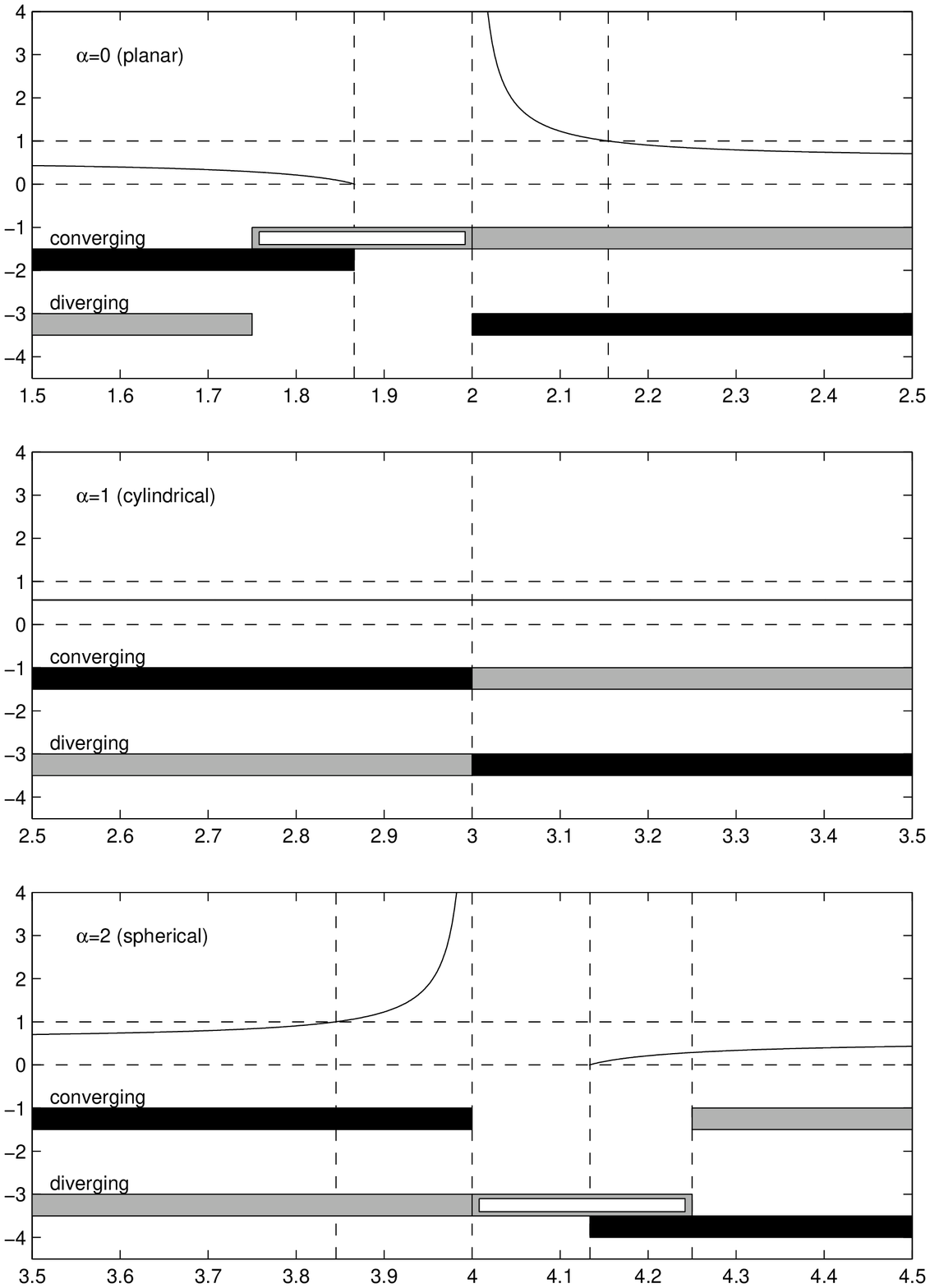}
\caption{\small {\it Schematic description of the possible solutions
for 1D strong relativistic explosions. The three panels are for planar
(top), cylindrical (middle) and spherical geometries (bottom).  The
$x$ axis is the density index $k$ ($\rho \propto r^{-k}$). Grey
stripes indicate the region where first type solutions exist ($k<4$ in
the spherical case), and black stripes show where a second type
solution exists. Hollow first type solutions, in which all the shocked
fluid is confined between the shock and some internal surface are
indicated by a hollow grey stripe. The y-axis of grey and black
stripes is arbitrary.  The solid line is the position of the sonic
point, $\chi_{sonic}$}.}
\label{fig:validity}
\end{figure}

A necessary condition for a second type solution to be correct, is
that its energy decrease with time, that is, $m>1+\alpha-k$ for
diverging solutions and the opposite for converging. We demonstrate
this now for diverging solutions in spherical geometry:
\begin{equation}
m_{II}=-2\alpha (5-3\sqrt{3})+(3-2\sqrt{3})k>1+\alpha-k=m_I
\end{equation}
or
\begin{equation}
k>(1+11\alpha-6\alpha\sqrt{3})/(4-2\sqrt{3})
\end{equation}
This gives $k>1+\sqrt{3/4}$ for planar symmetry, $k>3$ for cylindrical
symmetry and $k>5-\sqrt{3/4}$ for the spherical case. This is
completely consistent with the above requirements. Indeed the
requirement of decreasing energy follows from the condition that the
solution passes through the sonic point: if the flow behind the sonic
point cannot influence the flow ahead of the sonic point, energy
could not flow from behind the sonic.

\section{The Type-I to Type-II Transition}

Note, that when $m_{I}=m_{II}$ the two solutions listed above are the
same, despite their seemingly different expressions. Clearly this must
be the case since for a given $m$ the solution to the self similar
equations with the shock boundary conditions is unique. Nevertheless,
it is instructive to verify this from the two explicit expressions
given in the previous sections (e.g. equation \ref{eq:gfir} and
\ref{eq:gsec}). Let us consider the spherical case where $m_{I}=m_{II}$
at $k=5-\sqrt{3/4}$, and start by examining the second type
solution. For $k\approx 5-\sqrt{3/4}$, the numerator of equation
(\ref{eq:gsec}) is close to zero. This means than $g\chi$ has to be
close to unity to make the numerator close to zero as well. This
agrees with the expression for $g$ in the first type solution
$g=\chi^{-1}$.

Where is the sonic point in type-II solutions for $k$ close to
$k=5-\sqrt{3/4}$? Expanding equation (\ref{eq:gsec}) we obtain
\begin{equation}
g_{sonic} \rightarrow \left[ 1-\sqrt{3/4} \over k-(5-\sqrt{3/4}) \right]^{\sqrt{3/4}} 
\end{equation}
So that $g_{sonic}$ diverges when $k$ is close to $k=5-\sqrt{3/4}$.
Since at the sonic point $g_{sonic}\chi_{sonic}=4-2\sqrt{3}$ is
finite, $\chi_{sonic}$ approaches zero. This is very similar to the
behavior of Type-I solutions as those end at $\chi=0$ with divergence
of $g$. Nevertheless, Type-II solutions continue past $\chi=0$ into
negative $\chi$ and they are not hollow.

\begin{figure}[ht!]
\epsscale{0.75}
\plotone{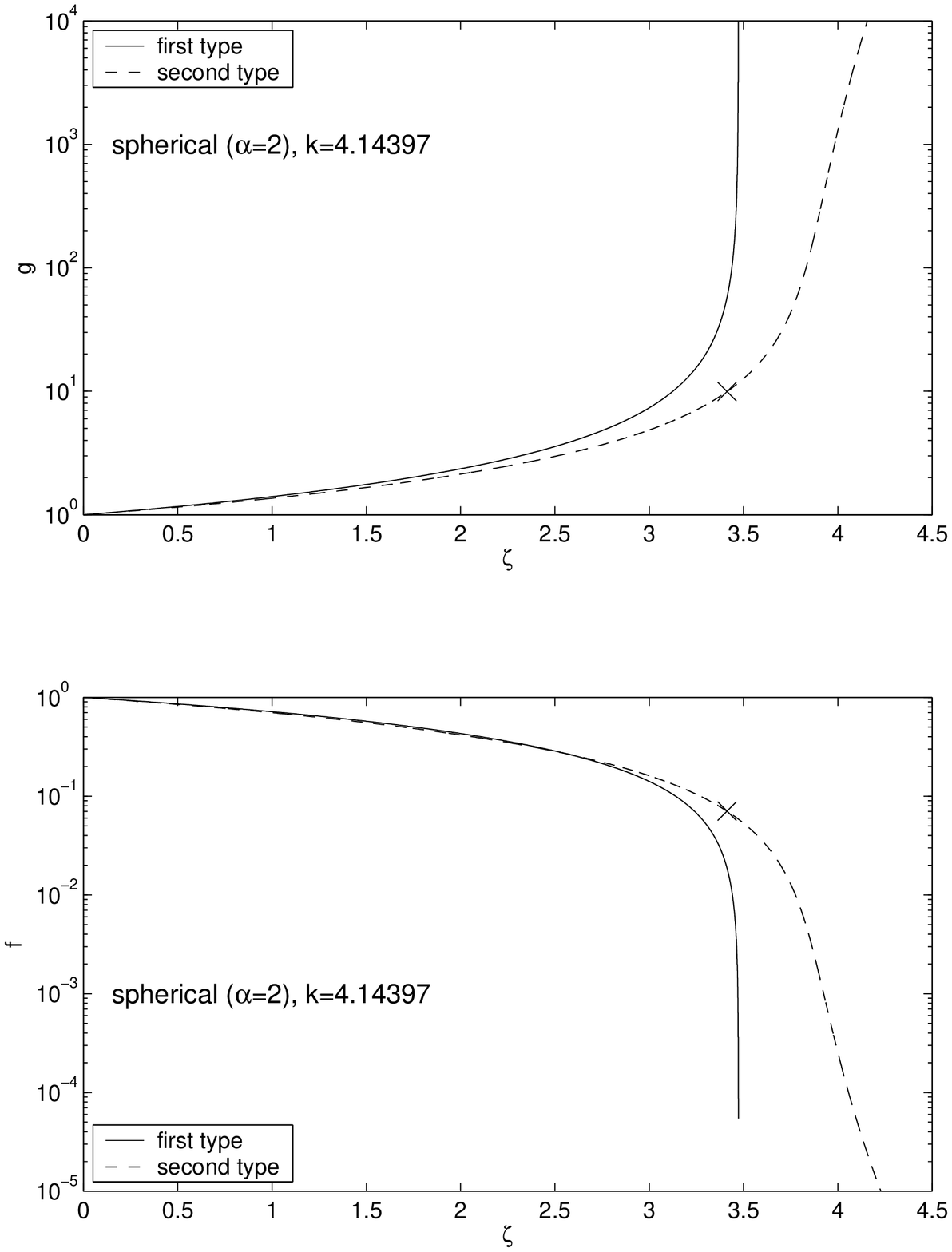}
\caption{\small {\it First and second type solutions for
$k=5-\sqrt{3/4}+0.01$ plotted as a function of $\zeta$.  The top plot
shows the Lorentz factor squared profile $g$ and the bottom plot the
pressure profile $f$.  The two solutions are very similar close to the
shock ($\zeta=0$). The Type-I solution is hollow, so the pressure
drops to zero and $\gamma$ diverges at some finite distance from the
shock $\zeta=-1/(2m+2) = 2+\sqrt{3}$.  The second type solution
continues smoothly but does show a relatively sharp drop around the
same distance.}}
\label{fig:transition}
\end{figure}

An unfortunate property of the solutions around $k=4$ (in spherical
geometry) is that the profile approaches non relativistic temperatures
very quickly as a function of the distance from the shock. We define
non relativistic temperatures to occur where $p/n < 1$. We have
\begin{equation}
{p\over n }\sim \Gamma {f\sqrt g \over h}  \sim \Gamma (1+2(4-k)\zeta)^{k+4 \over 6(k-4)} \sim \gamma \exp{(-{8\zeta/3})}
\end{equation}
So the temperature becomes non relativistic at
\begin{equation}
\zeta_{NR}\approx {3 \over 8} \ln \Gamma.
\end{equation}
This is just a few times $R/\Gamma^2$ behind the shock, even for quite
large $\Gamma$. For $k\approx 4$ the solution extends many times
$R/\gamma^2$ behind the shock. We therefore expect that for moderate
values of $\Gamma$, the ultra relativistic self similar solution gives
an approximate rather than an accurate description of the flow.

Clearly it is of importance to understand which of the two solutions
apply.  In principle, one can attempt this with numerical
simulations. However, the problem of non relativistic temperatures
noted above makes numerical investigation difficult as it requires
extremely large Lorentz factors which are hard to achieve.

\section{Discussion}

We have explored the possible ultra-relativistic self similar
solutions in planar, cylindrical, and spherical geometries, containing
diverging or converging shocks, and allowing for a general powerlaw
density profile $\rho \propto r^{-k}$. Relativistic implosions in all
three geometries and with a general powerlaw density profile are
treated here for the first time. While this paper was being prepared,
an independent paper appeared treating imploding relativistic shock
waves for spherical geometry with constant density \cite{hidalgo04}.

\begin{figure}[ht!]
\epsscale{0.6} \plotone{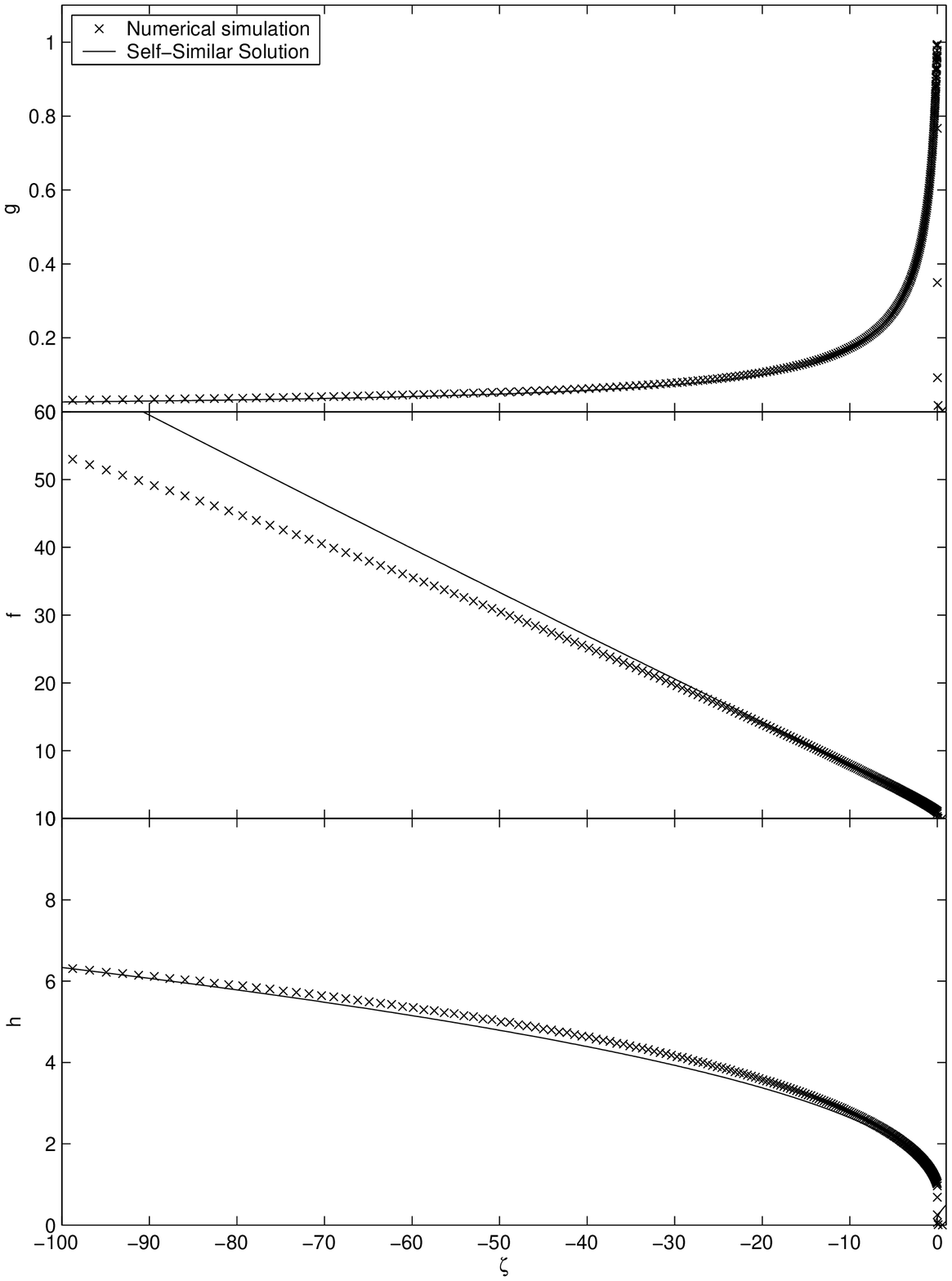}
\caption{\small {\it 
Comparison of numerical integration of the time dependent hydrodynamical
equations and the prediction of the self similar solution. Plotted,
top to bottom, are the Lorentz factor, the pressure and density
relative to that just behind the shock, for an imploding spherical 
shockwave into matter of constant density: $\alpha=2$, $k=0$.
}}
\label{fig:numerical}
\end{figure}

We show that as a function of the density powerlaw index $k$, there is
a sequence of self similar solutions. In the explosive scenarios, the
sequence is from first type solutions to hollow first type solutions
and then second type solutions as $k$ increases. The sequence is
reversed for imploding solutions. Those are of second type for low $k$
and first type for large $k$. In both cases, explosions and
implosions, the solution is of second type if the density ahead of the
shock decreases sufficiently fast and of first type if it increases
sufficiently fast. The physical interpretation of this is clear. A
shock wave that propagates into a density profile that decreases
sufficiently fast will accelerates and ``run away'' from the fluid
behind it. The shock becomes ``causally disconnected'' from the
downstream fluid, i.e. a second type solution. On the other hand, if
the shock wave does not accelerate enough, sound waves from behind it
will be able to catch up and the whole flow is causally connected,
i.e. a first type solution. In some of the geometries (converging
planar case and diverging spherical case) first type solutions become
hollow before they turn into second type solutions. This sequence of
self similar solutions and its physical origin apply to the non
relativistic case as well.

Our analysis raises some interesting riddles. In some cases (planar
diverging shocks or spherical converging shocks) this sequence has a
``gap'' between the first and second type solutions. In other cases
(planar converging shocks and spherical diverging shocks) there seems
to be an overlap in the sequence between first and second type
solutions: current considerations allow for the existence of both self
similar solutions for the same value of $k$.

Recently, Gruzinov \cite{gruzinov03} suggested a solution in the
``gap'' region in the non relativistic case which he called a ``third
type'' self similar solution. In his third type solution, the infinite
mass located at the origin acts like a piston moving at constant
velocity. Preliminary investigation seems to support a relativistic
analog of Gruzinov's analysis. In the relativistic case, however, the
solutions ara more interesting since the piston with large inertia
does not move at a constant speed but accelerates. Further research is
needed to establish this connection.

\acknowledgements
This research was partially supported by a NASA
ATP grant. RS is a Packard Fellow and an Alfred P. Sloan Research
Fellow. We thank Margaret Pan and Chris Matzner for helpful discussions.



\end{document}